\documentclass[aip,jcp,amsmath,amssymb,reprint,numerical,english,floatfix]{revtex4-1}
\usepackage{bm}
\usepackage{graphicx}

\begin{document}
\def\deriv#1#2{\frac{\mathrm{d}#1}{\mathrm{d}#2}}
\def\pderiv#1#2{\frac{\partial#1}{\partial#2}}
\def\e{\mathrm{e}}
\def\d{\mathrm{d}}
\def\vec#1{\bm{#1}}
\def\icm{\mathrm{\mathrm{cm}^{-1}}}
\def\imath{\mathrm{i}}
\def\ptderiv{\frac{\partial}{\partial t}}

\title{Vibrational vs. electronic coherences in 2D spectrum of molecular
systems}

\author{Vytautas Butkus}
\affiliation{Department of Theoretical Physics, Faculty of Physics, Vilnius
University, Sauletekio 9-III, LT-10222 Vilnius, Lithuania}
\affiliation{Center for Physical Sciences and Technology, Gostauto 9, LT-01108
Vilnius, Lithuania}

\author{Donatas Zigmantas}
\affiliation{Department of Chemical Physics, Lund University, P.O. Box
124, 22100 Lund, Sweden}

\author{Leonas Valkunas}
\affiliation{Department of Theoretical Physics, Faculty of Physics, Vilnius
University, Sauletekio 9-III, LT-10222 Vilnius, Lithuania}
\affiliation{Center for Physical Sciences and Technology, Gostauto 9, LT-01108
Vilnius, Lithuania}

\author{Darius Abramavicius}
\email{darius.abramavicius@ff.vu.lt}
\affiliation{Department of Theoretical Physics, Faculty of Physics, Vilnius
University, Sauletekio 9-III, LT-10222 Vilnius, Lithuania}
\affiliation{State Key Laboratory of Supramolecular Complexes, Jilin University,
2699 Qianjin Street, Changchun 130012, PR China}
\begin{abstract}
Two-dimensional spectroscopy has recently revealed the oscillatory
behavior of the excitation dynamics of molecular systems. However,
in the majority of cases there is considerable debate over what is
actually being observed: excitonic or vibrational wavepacket motion
or evidence of quantum transport. In this letter we present a method
for distinguishing between vibrational and excitonic wavepacket motion,
based on the phase and amplitude relationships of oscillations of
distinct peaks as revealed through a fundamental analysis of the two-dimensional
spectra of two representative systems.
\end{abstract}
\maketitle

Two-dimensional photon-echo (2DPE) spectroscopy is a powerful tool
capable of resolving quantum correlations on the femtosecond timescale
\cite{TianKeustersSuzakiEtAl2003,PhysRevLett.96.057406,engel-nat2007}.
They appear as  beats of specific peaks in the 2DPE spectrum for a
number of molecular systems \cite{Collini2009,engel-nat2007}. However,
the underlying processes are often ambiguous. At first, the beats
were attributed to the wave-like quantum transport with quantum coherences
being responsible for an ultra-efficient excitation transfer \cite{engel-nat2007,Collini2009,caruso-plenio-JCP09,Rebentrost-guzik-JPCB2009}.
The same process was associated with the opposite phase beats in the
spectral regions which are symmetric with respect to the diagonal
line \cite{Collini2010}. 

In molecules and their aggregates,  electronic transitions are coupled
to various intra- and intermolecular vibrational modes.  Vibrational
energies of these are of the order of 100 - 3000~cm$^{-1}$, while
the magnitudes of the resonant couplings, $J$, in excitonic aggregates
(e.g. in photosynthetic pigment-protein complexes or in J-aggregates)
are in the same range. Thus, vibronic and excitonic systems show considerable
spectroscopic similarities, and presence of electronic and/or vibrational
beats in the 2DPE spectrum is expected. Indeed, similar spectral beats
originating entirely from a high-energy vibrational wavepacket motion
have been observed \cite{Nemeth2010,Christensson2011}. The possibility
of distinguishing the electronic and vibrational origin of the beats
from  a 2DPE spectrum has been emphasized in a recent letter \cite{Turner2011JCPL}.
However, the reported conclusions  have not been supported by theoretical
arguments, and thus are questionable. Therefore, the highly relevant
question of how vibrations interfere with electronic coherences in
2DPE spectrum is still an open one. A theoretical study of the origin
of spectral beats, their phase relationships in the rephasing and
non-rephasing components of the 2DPE spectrum is presented in this
article.

We address this problem by considering two generic model systems which
exhibit distinct internal coherent dynamics. 
\begin{figure}
\noindent \begin{centering}
\emph{\includegraphics[width=7cm]{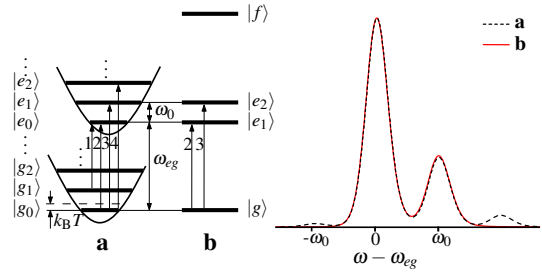}}
\par\end{centering}

\caption{\label{fig:Energy-level-structure}Energy level structure of the displaced
oscillator (a) and electronic dimer (b) and corresponding linear absorption
spectra.}
\end{figure}
The simplest model of an isolated molecular electronic excitation is
the vibronic system  represented by two electronic states, $|g\rangle$
and $|e\rangle$, which are coupled to a one-dimensional nuclear coordinate
$q$. We denote the model by a \emph{displaced oscillator} (DO) system
(Fig.~\ref{fig:Energy-level-structure}a). Taking $\hbar=$1, the
vibronic potential energy surface of the $|e\rangle$ state is shifted
up by electronic transition energy $\omega_{eg}$ and its minimum
is shifted by $d$ with respect to the ground state $|g\rangle$;
$d$ is the dimensionless displacement. This setup results in two
vibrational ladders of quantum sub-states $|g_{m}\rangle$ and $|e_{n}\rangle$,
$m,n=0\ldots\infty$, characterized by the Huang-Rhys (HR) factor
$HR=d^{2}/2$ \cite{MayBook2011,mukbook}. 

The other model system, which shows similar spectroscopic properties
but has completely different coherent internal dynamics without vibrations,
is an \emph{excitonic dimer} (ED). It consists of two two-level chromophores
(sites) with identical transition energies $\epsilon$. The two sites
are coupled by the inter-site resonance coupling $J$. As a result,
the ED has one ground state $|g\rangle$, two single-exciton states
$|e_{1}\rangle$ and $|e_{2}\rangle$ with energies $\varepsilon_{e_{1},e_{2}}=\epsilon\pm J$,
respectively, and a single double-exciton state $|f\rangle$ with
energy $\varepsilon_{f}=2\epsilon-\Delta$, where $\Delta$ is the
bi-exciton binding energy (Fig.~\ref{fig:Energy-level-structure}b)\cite{valkunasbook}.

The absorption spectrum of both systems is as follows. The absorption
of the  DO is determined by transitions from the $|g_{m}\rangle$
vibrational ladder into $|e_{n}\rangle$ scaled by the Franck-Condon
(FC) vibrational wavefunction overlaps \cite{MayBook2011,valkunasbook}.
Choosing $HR=0.3$ and $k_{\mathrm{B}}T\approx\frac{1}{3}\omega_{0}$
and assuming Lorentzian lineshapes with linewidth $\gamma$, we get
the vibrational progression in the absorption spectrum (dashed line
in Fig.~\ref{fig:Energy-level-structure}). Here $\omega_{0}$ is
the vibrational energy. The most intensive peaks at $\omega_{eg}$
and $\omega_{eg}+\omega_{0}$ correspond to 0-0 and 0-1 vibronic transitions.
Qualitatively similar peak structure is featured in the absorption
of ED, where the spectrum shows two optical transitions $|g\rangle\to|e_{1}\rangle$
and $|g\rangle\to|e_{2}\rangle$, assuming both are allowed. Choosing
$J=\omega_{0}/2$ and the angle $\varphi$ between the chromophore
transition dipoles equal to $\pi/6$, and using adequate linewidth
parameters, we get absorption peaks (solid line in Fig.~\ref{fig:Energy-level-structure})
that exactly match the strongest peaks of the DO. As expected one
cannot distinguish between these two internally different systems
from the absorption spectra alone.

The 2DPE spectrum carries more information than absorption. However,
it consists of many contributions and unambiguous distinction between
the ED and DO systems becomes difficult. In order to unravel the 2DPE
spectra we thus need to construct the entire 2D signal from the first
principles for both systems and recover the source of oscillations
in the 2DPE spectrum. 

In the conventional scheme of the 2DPE measurement, two primary excitation
pulses with wavevectors $\bm{k}_{1}$ and $\bm{k}_{2}$ followed by
the probe pulse $\bm{k}_{3}$ are used; $\bm{k}_{j}$ are pulse wavevectors.
The signal is detected at the $\bm{k}_{\mathrm{S}}=-\bm{k}_{1}+\bm{k}_{2}+\bm{k}_{3}$
phase-matching direction. The order of $\bm{k}_{1}$ and $\bm{k}_{2}$
defines the rephasing configuration ($\vec k_{\mathrm{I}}$) when
$\bm{k}_{1}$ comes first and the non-rephasing configuration ($\vec k_{\mathrm{II}}$)
when $\bm{k}_{2}$ comes first. 

\begin{figure}
\begin{centering}
\includegraphics[width=7cm]{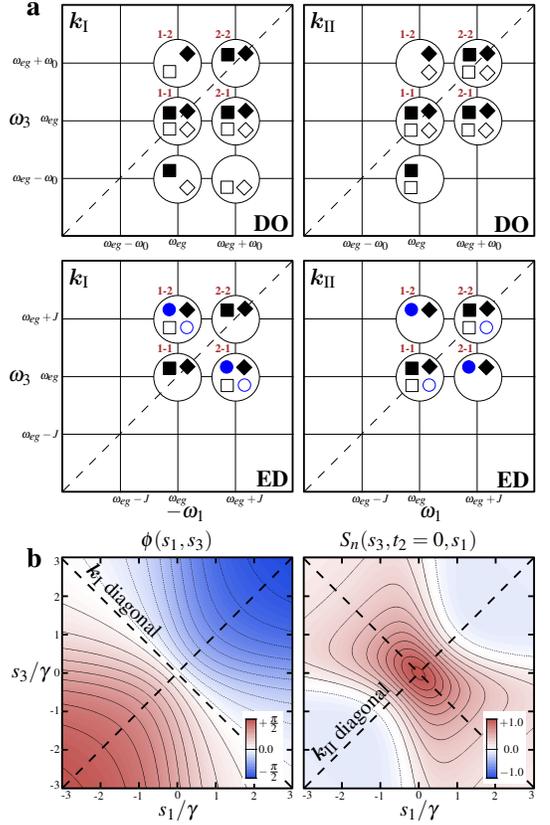}
\par\end{centering}

\caption{\label{fig:Double-sided}(a) Scheme of contributions to 2DPE spectrum
of the $\bm{k}_{\mathrm{I}}$ and $\bm{k}_{\mathrm{II}}$ signals
for the reduced DO and ED ($\Delta=0$) systems. The ESE contribution
is indicated by squares, GSB -- diamonds, ESA -- circles. Solid symbols
denote non-oscillating contributions in $t_{2}$, open -- oscillatory
in the form of $\pm\cos(\varepsilon_{2}t_{2})$, where $\varepsilon_{2}=\omega_{0}$
for DO and $\varepsilon_{2}=2J$ for ED. (b) Phase $\phi$ of the
contribution (Eq. \ref{eq:sig-all}) and peak profile $S_{n}$ as
a function of the shift from the peak center ($s_{1}=s_{3}=0$) using
relative coordinates. The diagonal lines of the $\vec k_{\mathrm{I}}$
and $\vec k_{\mathrm{II}}$ contributions to the 2D spectra are shown
by dashed lines. Peaks are labeled in plots as `1-1', `1-2', etc.}
\end{figure}
Semiclassical perturbation theory with respect to the incoming fields
reveals the system-field interaction and evolution sequences, often
denoted by the Liouville space pathways. Three types of distinct interaction
configurations are denoted by the Excited State Emission (ESE), Ground
State Bleaching (GSB) and Excited State Absorption (ESA) contributions
\cite{Abramavicius-EPL2007,mukbook}. If we neglect environment-induced
relaxation, the signals are given as sums of resonant contributions,
$S(\omega_{3},t_{2},\omega_{1})=\sum_{n}S_{n}(\omega_{3},t_{2},\omega_{1})$
of the type 
\begin{eqnarray}
S_{n}(\omega_{3},t_{2},\omega_{1}) & = & A_{(n)}\iint\mathrm{d}t_{1}\mathrm{d}t_{3}\mathrm{e}^{+\mathrm{i}\omega_{3}t_{3}+\mathrm{i}\omega_{1}t_{1}}\nonumber \\
 &  & \times[\pm G_{3}(t_{3})G_{2}(t_{2})G_{1}(t_{1})]_{(n)},\label{eq:respfun}
\end{eqnarray}
where the subscript $n$ denotes different terms of the summation.
$A_{(n)}$ is a complex prefactor, given by the transition dipoles
and excitation fields, the propagator of the density matrix $G$ for
the $j$th ($j=1,2,3)$ time delay is of the one-sided exponential
function type 
\begin{equation}
G_{j}(t_{j})=\theta(t_{j})\exp(-\mathrm{i}\varepsilon_{j}t_{j})
\end{equation}
 ($\theta(t)$ is the Heaviside step-function). Here $\varepsilon_{j}$
coincides with the energy gap $\omega_{ab}$
between the \emph{left} and \emph{right }states of the system density
matrix relevant to the time interval $t_{j}$. ESE and GSB carry `$+$'
sign while ESA has `$-$' overall sign. 

The Fourier transforms in Eq.~\eqref{eq:respfun} map the contributions
to the frequency-frequency plot $(t_{1},t_{3})\to(\omega_{1},\omega_{3})\sim(\mp|\varepsilon_{1}|,\varepsilon_{3})$
(the upper sign is for $\vec k_{\mathrm{I}}$, the lower -- for $\vec k_{\mathrm{II}}$).
Diagonal peaks at $\omega_{1}=\mp\omega_{3}$ are usually distinguished,
while the anti-diagonal line is defined as $\mp\omega_{1}+\omega_{3}=Const$.
The whole 2DPE signal becomes a function of $t_{2}$: either oscillatory
for density matrix coherences $|a\rangle\langle b|$ with characteristic
oscillation energy $\varepsilon_{2}=\omega_{ab}\neq0$, or static
for populations $|a\rangle\langle a|$ ($\varepsilon_{2}=0$). 

To reveal oscillatory contributions in the DO and ED systems we have
grouped all contributions into either oscillatory or static as
shown in Fig.~\ref{fig:Double-sided}a. If we consider only the two
main vibrational sub-states in DO, the 2DPE spectrum will have only
ESE and GSB contributions, while ED additionally has ESA. As a function
of $t_{2}$, the DO system has 8 oscillatory and 8 static configurations,
which organize into six peaks, while the ED system has only 4 oscillatory
and 8 static contributions which give four peaks. The net result is
that the diagonal peaks in the $\vec k_{\mathrm{I}}$ and cross-peaks
in the $\vec k_{\mathrm{II}}$ signals are non-oscillatory in the
ED, while all peaks except the upper diagonal peak in $\vec k_{\mathrm{I}}$
are oscillatory in the DO. We thus find significant differences in
oscillatory peaks between ED and DO systems.

An important additional parameter to consider is a phase of oscillation.
Eq.~\eqref{eq:respfun} can be analytically integrated for $G_{j}(t_{j})\propto\exp(-\mathrm{i}\varepsilon_{j}t_{j}-\gamma_{j}t_{j})$.
For a single contribution $S_{n}$ giving rise to a peak at $(\omega_{1},\omega_{3})=(\mp|\varepsilon_{1}|,\varepsilon_{3})$
we shift the origin of $(\omega_{1},\omega_{3})$ plot  to the peak
center by introducing the displacements ($\omega_{1}+\varepsilon_{1}=-s_{1}$,
$\omega_{3}-\varepsilon_{3}=s_{3}$ for the rephasing pathways, while
$\omega_{1}-\varepsilon_{1}=s_{1}$, $\omega_{3}-\varepsilon_{3}=s_{3}$
for the nonrephasing). For $\gamma\approx\gamma_{1}\approx\gamma_{3}$
we get the peak profile
\begin{equation}
S_{n}(s_{3},t_{2},s_{1})=A_{n}L(s_{1},s_{3})\mathrm{e}^{-\gamma_{2}t_{2}}\cos(\left|\varepsilon_{2}\right|t_{2}+\phi(s_{1},s_{3})),\label{eq:sig-all}
\end{equation}
 where the lineshape and phase  for the $\vec k_{\mathrm{I}}$ (upper
sign) and $\vec k_{\mathrm{II}}$ (lower sign) signals are  
\begin{eqnarray}
L(s_{1},s_{3}) & = & \frac{\sqrt{[\gamma^{2}\pm s_{1}s_{3}]^{2}+\gamma^{2}(s_{3}\mp s_{1})^{2}}}{(s_{1}^{2}+\gamma^{2})(s_{3}^{2}+\gamma^{2})},\\
\phi(s_{1},s_{3}) & = & \mathrm{sgn}\left(\varepsilon_{2}\right)\arctan\left(\frac{\gamma(s_{3}\mp s_{1})}{\left(\mp s_{1}s_{3}-\gamma^{2}\right)}\right).
\end{eqnarray}
 The phase $\phi$ and the full profile for $A_{n}=1$ and $t_{2}=0$ are
shown in Fig.~\ref{fig:Double-sided}c. The rephasing and non-rephasing
configurations are obtained by flipping the direction of the $s_{1}$
axis. At the center of the peak ($s_{1}=s_{3}=0$), we have and $\phi=0$,
leading to $S_{n}\propto\cos(\left|\varepsilon_{2}\right|t_{2})$.
However, for ($s_{1}\neq0,s_{3}\neq0$) we find $S_{n}\propto\cos(\left|\varepsilon_{2}\right|t_{2}+\phi(s_{1},s_{3}))$
with $\phi(s_{1},s_{3})\neq0$.\emph{ }Thus,\emph{ the displacement
from the peak center determines the phase of the spectral oscillations}.
Note that the sign of the phase $\phi$ is opposite for the peaks
above ($\varepsilon_{2}<0$) and below ($\varepsilon_{2}>0$) the
diagonal line, and this applies for all contributions.

The whole 2DPE spectrum is a sum of all relevant contributions. Assuming
that all dephasings are similar, different contributions to the same
peak will have the same spectral shape and they may be summed. We
can then simplify the 2DPE plot by writing the signal as a sum of
peaks $\bar{\sum}$, which have static (from populations) and oscillatory
(from coherences) parts: 
\begin{eqnarray}
 &  & S(\omega_{3},t_{2},\omega_{1})=\mathrm{e}^{-\gamma_{2}t_{2}}\bar{\sum}_{i,j}L_{ij}(\omega_{1},\omega_{3})\nonumber \\
 &  & \quad\times\left[A_{ij}^{\mathrm{p}}+A_{ij}^{\mathrm{c}}\cdot\cos(|\omega_{ij}|t_{2}+\phi_{ij}(\omega_{1},\omega_{3}))\right].\label{eq:peaks}
\end{eqnarray}
Here $\omega_{ij}$ is the characteristic oscillatory frequency of
a peak ($ij$), $A_{ij}^{\mathrm{p}}(t_{2})$ and $A_{ij}^{\mathrm{c}}(t_{2})$
are the real parts of orientationally-averaged prefactors of population
and coherence (electronic or vibronic) contributions, respectively.
The spectral lineshape is given by $L_{ij}(\omega_{1},\omega_{3})$.
Here we clearly identify the oscillatory amplitude and its phase for
a specific peak. 

To apply this expression to our systems, we assume a typical situation
where the spectrum of the laser pulses is tuned to the center of the
absorption spectrum and the limited bandwidth selects the two strongest
absorption peaks. In the 2DPE spectra two diagonal and two off-diagonal
peaks for ED and DO are observed. Indices $i$ and $j$ in Eq.~\eqref{eq:peaks}
run over the positions of the peaks and thus can be (1,1), (1,2),
(2,1), and (2,2). For clarity we study the spectral dynamics with
$t_{2}$ at the short delays, $t_{2}\ll\gamma_{2}^{-1}$, and use
notations $A,\, L$ for the $\vec k_{\mathrm{I}}$ signal and $\tilde{A},\,\tilde{L}$
for the $\vec k_{\mathrm{II}}$ signal. 

The transition dipole properties of the ED results in the picture
where all static amplitudes of the ED are positive and $A_{11}^{\mathrm{p}}=\tilde{A}_{11}^{\mathrm{p}}$,
$A_{22}^{\mathrm{p}}=\tilde{A}_{22}^{\mathrm{p}}$, $A_{12}^{\mathrm{p}}=A_{21}^{\mathrm{p}}=\tilde{A}_{21}^{\mathrm{p}}=\tilde{A}_{12}^{\mathrm{p}}$.
The oscillatory amplitudes are equal: $A_{12}^{\mathrm{c}}=A_{21}^{\mathrm{c}}=\tilde{A}_{11}^{\mathrm{c}}=\tilde{A}_{22}^{\mathrm{c}}$. Such relationships are obtained by considering the all-parallel organization of polarization of incoming electric fields and neglecting the bi-exciton binding energy. The spectral beats with $t_{2}$ can thus only have the same phases
in the $\vec k_{\mathrm{I}}$ or $\vec k_{\mathrm{II}}$ spectrum,
when measured at peak centers. Additionally the oscillatory ESE and
ESA parts in ED cancel each other if $\Delta=0$ and their broadenings
are equal. As these relationships do not depend on coupling $J$ and
transition dipole orientations, all ED systems should behave similarly.
By studying the whole parameter space, it can also be shown  that
these relations hold for a hetero-dimer.

The amplitude-relationships, however, are different for the DO system.
\begin{figure}
\begin{centering}
\includegraphics[width=7cm]{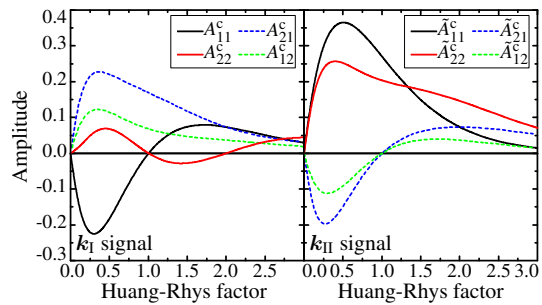}
\par\end{centering}

\caption{\label{fig:Amplitudes_monomer}The amplitudes of oscillatory peaks
of 2D spectra of the DO model for $\vec k_{\mathrm{I}}$ and $\vec k_{\mathrm{II}}$
signal. Note that the negative amplitude denotes a phase shift of
$\pi$of the oscillation.}
\end{figure}
 The amplitudes $A_{ij}^{(c)}$ of the oscillatory peaks are plotted
in Fig.~\ref{fig:Amplitudes_monomer} as a function of the HR factor,
where now we include all vibrational levels in the $|g_{m}\rangle$
and $|e_{n}\rangle$ ladders. For $\vec k_{\mathrm{I}}$, the amplitudes
$A_{11}^{\mathrm{c}}$ and $A_{22}^{\mathrm{c}}$ maintain the opposite
sign when $HR<2$ and are both positive when $2<HR<3$ (note that
$A_{22}^{\mathrm{c}}=0$ when only two vibrational levels are considered
in Fig.~\ref{fig:Double-sided}b). The oscillation amplitudes $A_{11}^{\mathrm{c}}$
and $A_{22}^{\mathrm{c}}$ change sign at $HR=1$. Amplitudes $A_{12}^{\mathrm{c}}$
and $A_{21}^{\mathrm{c}}$ are always positive. Spectrum oscillations
with $t_{2}$ for both diagonal peaks in the $\bm{k}_{\mathrm{II}}$
signal will be in-phase for the whole range of the HR factor. The
same pattern holds for the 1-2 and 2-1 cross-peaks, which will oscillate
in-phase, but will be of opposite phase compared to the diagonal peaks
in the region of $HR<1$. Note that the sign of amplitudes changes
with the HR factor, since the overlap integral between vibrational
wavefunctions can be both positive and negative. The amplitudes of
static contributions are positive in the whole range of parameters
and are identical for both $\vec k_{\mathrm{I}}$ and $\vec k_{\mathrm{II}}$
signals. 

We thus find very different behaviour of oscillatory peaks of DO and
ED systems. The above analysis applies for the central positions of
the peaks, which may be difficult to determine if the broadening is
large. Note that the phase $\phi$ varies from $-\pi/2$ to $+\pi/2$
(Eq.~\eqref{eq:sig-all} and Fig.~\ref{fig:Double-sided}c) when
probing in the vicinity of the peak. However, $\phi=0$ along the
diagonal line for $\vec k_{\mathrm{I}}$ and along the anti-diagonal
line for $\vec k_{\mathrm{II}}$. These lines can thus be used as
guidelines for reading phase relations of distinct peaks in the 2DPE
spectrum. For instance, the two diagonal peaks can be calibrated by
reading their amplitudes at the diagonal line, or the two opposite
cross-peaks can be compared by drawing anti-diagonal lines.

\begin{figure}
\begin{centering}
\includegraphics[width=7cm]{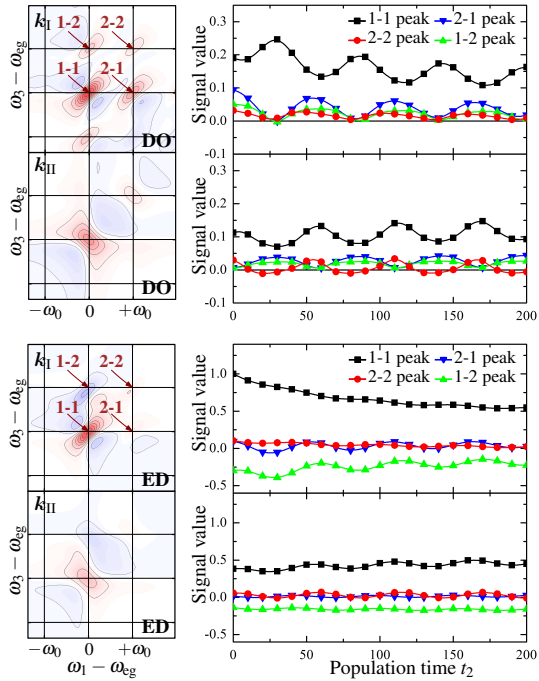}
\par\end{centering}

\caption{\label{fig:Intensities-of-peaks}2DPE spectra and peak values of the
DO and ED as the function of population time $t_{2}$, of the $\vec k_{\mathrm{I}}$
and $\vec k_{\mathrm{II}}$ signals. Spectra are normalized to the
maxima of the total spectra of the DO and ED.}
\end{figure}

The 2DPE spectra for both DO and ED systems calculated by including
phenomenological relaxation and Gaussian laser pulse shapes ~\cite{AbramaviciusValkunasCP2010,Abramavicius-EPL2007}
are plotted in Fig.~\ref{fig:Intensities-of-peaks}.
 The structure and the $t_{2}$ evolution of the spectra illustrate
the dynamics discussed above and clearly shows the distinctive spectral
properties of the vibronic vs. electronic system: (i) diagonal peaks
in the $\vec k_{\mathrm{I}}$ signal are oscillating in DO, but only
exponentially decaying in ED (the oscillatory traces come from the
overlapping tails of off-diagonal peaks), (ii) the relative amplitude
of oscillations is much stronger in DO as compared to ED, where the
ESA and ESE cancellation suppresses the oscillations, (iii) opposite
oscillation phases are observed in DO, while all peaks oscillate in-phase
in ED. 

The up-to-date experiments are capable of creating broad-band pulses
\cite{Christensson2011}. Thus, the overtones in DO can be excited
and beats of $n\omega_{0}$ frequencies ($n$ is integer) observed.
These may become important in the case of large HR factors. Such frequencies
are absent in the ED system, since only one oscillatory frequency,
equal to $2J$ is available.

The analysis presented in this article provides a clear physical picture
of electronic and vibronic coherence beatings in 2DPE spectra. We
are able to discriminate weakly damped electronic and vibronic coherent
wavepackets in molecular systems based on fundamental theoretical
considerations. Dynamics of diagonal peaks and cross-peaks as well
as relative phase between them in the rephasing signal can now be
classified for vibrational and excitonic systems as follows. (i) Static
diagonal peaks and oscillatory off-diagonal peaks signify pure electronic
coherences, not involved in energy transport. (ii) Oscillatory diagonal
peaks in accord with off-diagonal peaks ($0$ or $\pi$ phase relationships)
signify vibronic origins. The oscillation phase is $0$ for electronic
coherences and $0$ or $\pi$ for vibronic coherences. These outcomes
hold if\emph{ }the signal is probed at the very centers of the spectral
resonances. The observed phase of the beatings varies as the signal
is recorded away from the center of an oscillating peak. 

Our results might be useful in analysis of recently observed beatings
in molecular systems. For instance phase relations of the beatings
detected at separate points in the vicinity of the same cross-peak
of the photosynthetic LH2 complex \cite{Harel03012012} might be 
the result of measurement away from the peak center (see Fig.~\ref{fig:Double-sided}b).
The issue of probing away from the peak centers also applies to the
opposite-phase beatings reported by Collini et al. \cite{Collini2010}.
Our analysis thus shows that the detailed phase relationships in the
two dimensional spectra may be of critical importance. By helping
to identify spectral beats in photosynthetic aggregates, the presented
analysis should facilitate answering the question of importance of
electronic coherences in excitonic energy transfer, its efficiency
and robustness. 

This research was funded by the European Social Fund under the Global
Grant measure.


\end{document}